\documentclass{revtex4}
\usepackage[utf8]{inputenc}  
\usepackage{amsmath}
\usepackage{amssymb}
\usepackage{graphicx}
\usepackage{epstopdf}

\begin{document}
    
\title{Effect of edge vacancies on performance of planar graphene tunnel field-effect transistor}

\author{Glebov A.A$^{1,2}$, Katkov V.L$^{1}$, Osipov V.A.$^{1}$}
\affiliation{$^{1}$Bogoliubov Laboratory of Theoretical Physics, Joint Institute for Nuclear Research, 141980 Dubna, Moscow region, Russia}
\affiliation{$^{2}$Moscow Institute of Physics and Technology, 141700 Dolgoprudny, Moscow region, Russia}

\begin{abstract}
    
The influence of edge vacancies on the working ability of the planar graphene tunnel field-effect transistor (TFET) is studied at various concentrations and distributions (normal, uniform, periodic) of defects. All calculations are performed by using the Green's function method and the tight-binding approximation. It is shown that the transistor performance depends critically on two important factors associated with the defects: the destruction of the edge-localized electronic states and the emergence of subpeaks near the Fermi level. The supportable operation conditions of the TFET are found to be ensured at 30 percent or less of edge vacancies regardless of the type of their distribution. 

\end{abstract}

\maketitle

\section{Introduction}

Graphene layers with zigzag edges have a broad potential application as the building blocks for various nanoelectronic and spintronic devices~\cite{schwierz,heerema,bergvall,sadeghi,isaeva}. 
Recently, a concept of the planar graphene tunnel-field effect transistor (TFET) has been proposed~\cite{katkov, katkov2}. The main idea was based on the use of specific edge state effects in graphene electrodes with zigzag termination in the regime of tunnel current.
The suggested TFET has a rather simple configuration consisting of two facing each other in-plane electrodes with a nanogap between them. 
While tunneling has not been measured so far, current progress in STM lithography shows the feasibility for the fabrication of such device in the near future~\cite{he, baringhaus}.

The zigzag edges of graphene host edge-localized states which are characterized by a high electronic density at the Fermi level~\cite{klein, nakada, fujita}. 
These states were experimentally observed using tunneling spectroscopy and STM~\cite{ritter, tao}. 
What is important, the edge states are “topologically protected” in the sense that their existence is based on a topological property (a non-trivial Zack phase) and remain robust to weak external perturbations~\cite{deplace}. 
This feature ensures the stable performance of the planar graphene TFET.

At the same time, graphene electrodes can accommodate different kinds of defects. 
The most important of them are edge vacancies that can change the density of electronic states (DOS) at the edges. 
As was recently shown, the very existence of the edge states depends on both concentration and distribution of defects~\cite{glebov}. 
Obviously, vacancies will affect the tunnel current through the edge states which are the working basis of the TFET. 
The aim of this work is to study the influence of both the concentration and different distribution (normal, uniform, periodic) of vacancy defects on the operation of the planar graphene tunnel field-effect transistor.

\section{Model}

\begin{figure}[h]
  \includegraphics{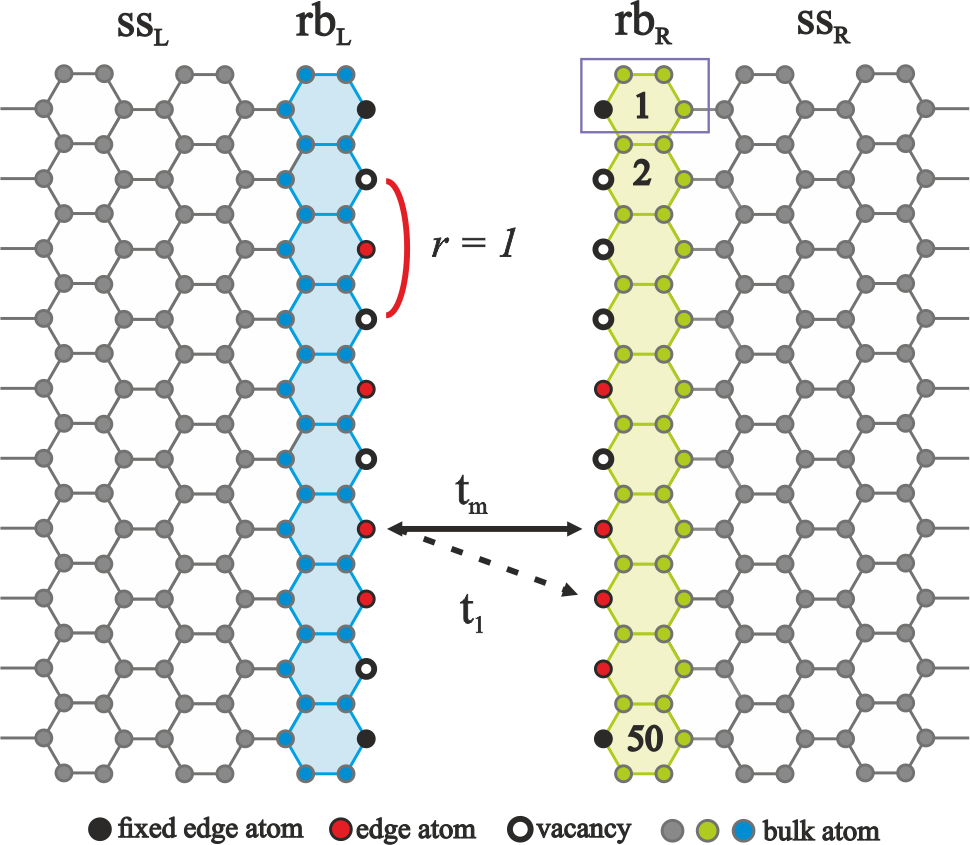}
\caption{Graphene lattice divided into two blocks: semi-infinite (in the horizontal direction) sheet $ss_{R,L}$ and graphene ribbon $rb_{R,L}$. Colored  cells are numbered from 1 to 50. $t_{m}$ and $t_{1}$ are the hopping parameters between the contacts and $r$ defines the number of atoms between vacancies.}
\label{gk}
\end{figure}

The atomistic structure of TFET contacts is shown in fig.\ref{gk}.
It is convenient to divide each graphene contact into two parts: a semi-infinite sheet ($ss_{R,L}$) and a ribbon ($rb_{R,L}$) containing the vacancies. 
The ribbon is taken to comprise of $L=50$ cells. Actually, we considered different values of $L$ ($L=100$, 400) and found that the results are insensitive to the ribbon width.
Vacancies may appear on any positions of edge atoms except 1 and 50.
This structure is described by the Hamiltonian consisting of two parts: $H_{R,L}^{ss}$ and $H_{R,L}^{rb}$.
Vacancy defects are taken into account by replacement of zero value in the diagonal cell of the matrix $H^{rb}_{R,L}$ for the corresponding edge atom to infinity~\cite{li}.
The calculations are performed by using the Green's function method and the tight-binding approximation.
The edge Green's function for the isolated graphene contact is given by
\begin{equation}
    g_{R,L}^{-}(\epsilon)=[(\epsilon+i0^{+})\textbf{1}-H^{rb}_{R,L}-Tg^{ss}_{R,L}T^{\dagger}]^{-1},
\end{equation}
where
\begin{equation}
    g^{ss}_{R,L}(\epsilon)=[(\epsilon+i0^{+})\textbf{1}-H^{ss}_{R,L}]^{-1}.
\end{equation}
Here $\epsilon$ is the energy, $T$ is the interaction matrix between the $ss$ and $rb$, 
$g^{ss}_{R,L}$ is the retarded Green's function for the sheet,
$\textbf{1}$ is the identity matrix. 
$g^{ss}_{R,L}$ is calculated by using of the iterative algorithm described in refs.~\cite{sancho, sancho2}.
\begin{figure}[h]
\includegraphics{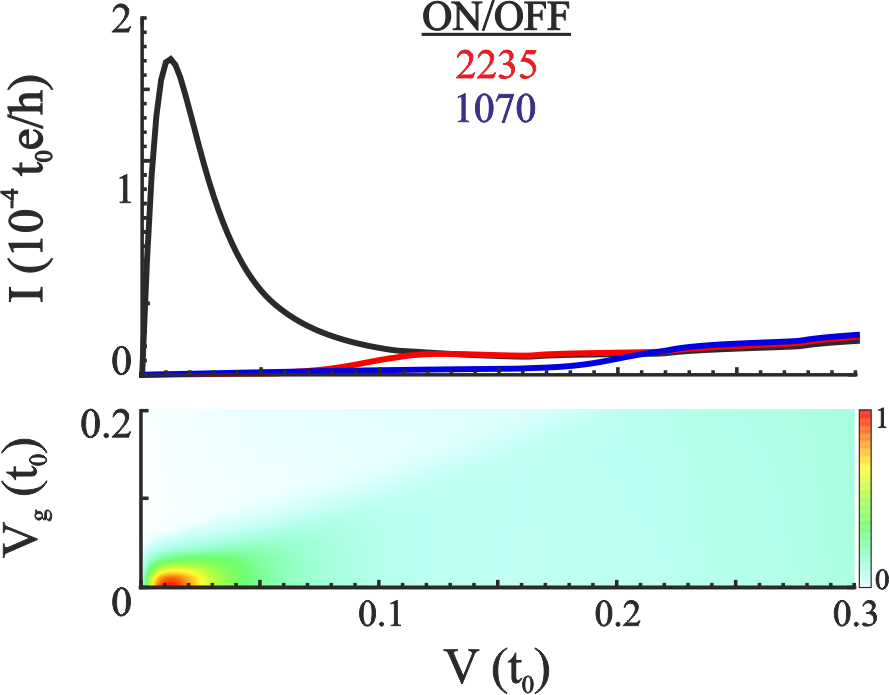}
\caption{I-V characteristics of the planar graphene TFET with defect-free contacts at different gate voltages $V_g$ (in $V$). The gate voltage is taken to be $V_{g}=0$ (black), $0.1$ (red), and $0.2$ (blue). The ON/OFF ($(I_{on}/I_{off})^{max}$) ratios of the device are shown. Colour map shows current density normalized to the highest value for a range of $V$ and $Vg$.} 
\label{id}
\end{figure}
The tunnel current is written as~\cite{berthod, todorov}
\begin{equation}
I=\frac{e}{h}\int\limits_{-\infty}^{+\infty} \mathrm{Tr}[A_{L}(\mathbf{1}-t^{\dagger}g_{R}^{-}tg_{L}^{-})^{-1}        t^{\dagger}A_{R}t(\mathbf{1}-g_{L}^{+}t^{\dagger}g_{R}^{+}t)^{-1}]\times[f_{L}-f_{R}]d\epsilon,
\end{equation}  
where 
$g_{R,L}^{-}=(g_{R,L}^{+})^{\dagger}$ is the retarded Green's function for the right (left) contact containing the vacancies,
$A_{L,R}=i(g_{R,L}^{+}-g_{R,L}^{-})$ is the spectral density, 
$t$ is the interaction matrix between the left and the right contacts, 
$f_{L,R}=f(\epsilon\mp eV/2)$,
$f(\epsilon)=[1+\textsl{exp}(\epsilon/k_{B}T)]^{-1}$ is the Fermi distribution function, $k_{B}T=0.01t_{0}$, $t_{0}$ is taken to be 2.86 eV.
The tunnel current is normalized to the cell containing four atoms (see the violet rectangle in fig.~\ref{gk}).
The influence of gate voltage on both contacts is counted by the replacement $H_{R,L}(\epsilon)\rightarrow H_{R,L}(\epsilon+V_{g})$. 

The tunneling parameters ($t_{m}$ and $t_{1}$ in fig.~\ref{gk}) are calculated using the known relation ~\cite{bardeen, duke}
\begin{equation}
t(r_{0})=\frac{-h^{2}}{2m}\int\limits_{S}[p_{z}^{*}(r)\nabla p_{z}(r-r_{0})-p_{z}(r-r_{0})\nabla p_{z}^{*}(r)]\textit{d}^{2}r,
\end{equation}   
 where $r_{0}$ is a distance between carbon atoms and $S$ is a surface between their $p_{z}$ orbitals whose values are taken from the Herman-Skillman tables~\cite{herman}.
Our analysis shows that at distances greater than $2.5a_{0}$ the values of $t(r_{0})$ can be approximated with good accuracy by 
\begin{equation}
\label{approx}
    t(d)=0.7t_{0}\exp[-2.2(d/a_{0}-1)],
\end{equation}
where $d$ is a distance between atoms in different contacts and $a_{0}=0.142$ nm. We obtain that $t_{m}=0.009t_{0}$ at $d=0.426$ nm and $t_{1}=0.002t_{0}$ at $d=0.492$ nm.   
Notice that a similar to eq.~(\ref{approx}) expression was used in the description of the charge transfer between atoms in a pair of DNA nucleotides~\cite{hawke}.

\section{Effect of vacancies on the tunnel current}

\begin{figure*}[!pt]
\includegraphics[scale=0.9]{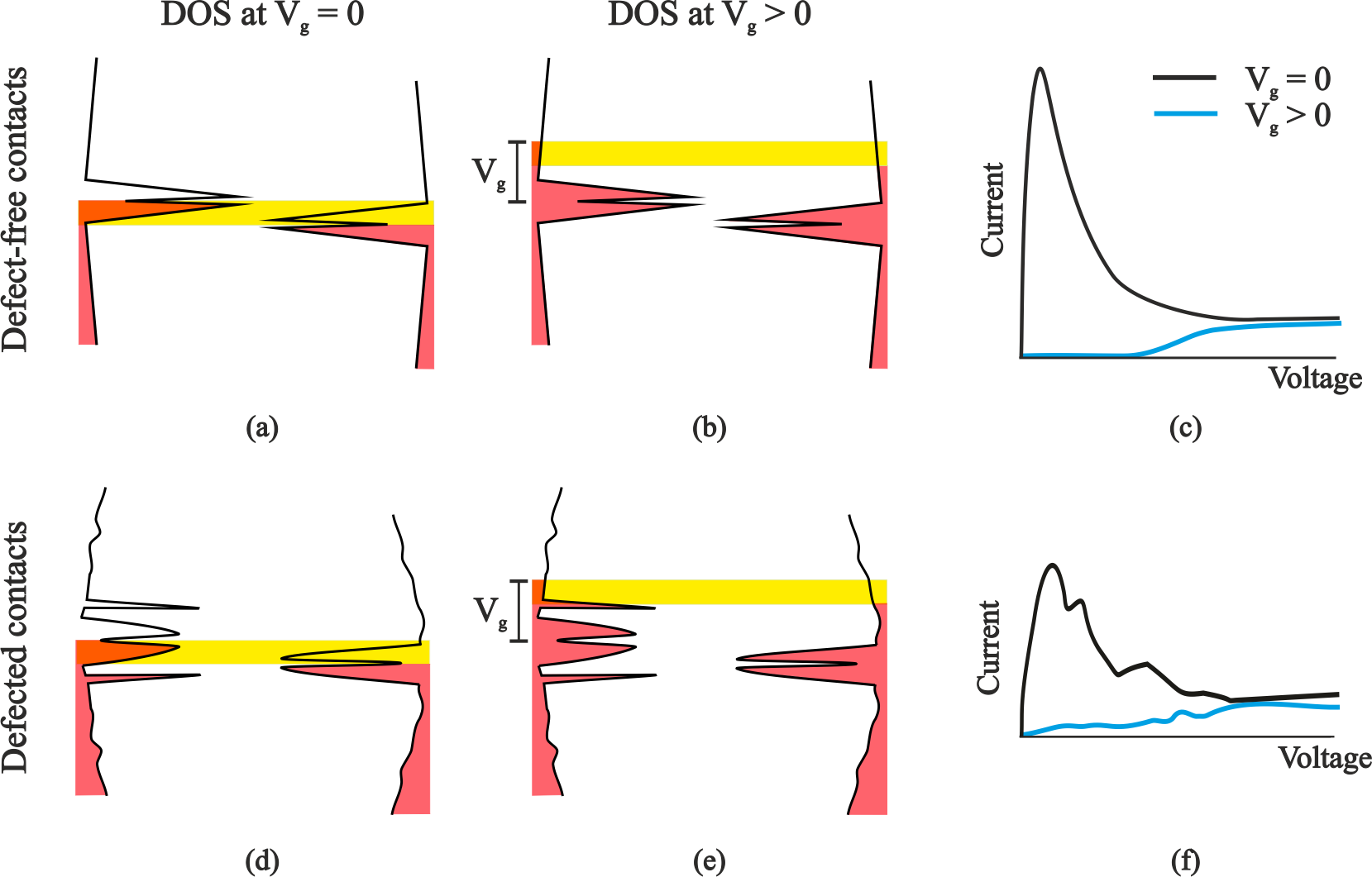}
\caption{Schematic electronic total DOS diagrams (left) and the corresponding I-V curves of the TFET (right) for both defect-free (a,b,c) and defected graphene contacts (d,e,f). Filled states are painted in red, yellow rectangles show the energy window.} 
\label{dos}

\includegraphics[scale=0.65]{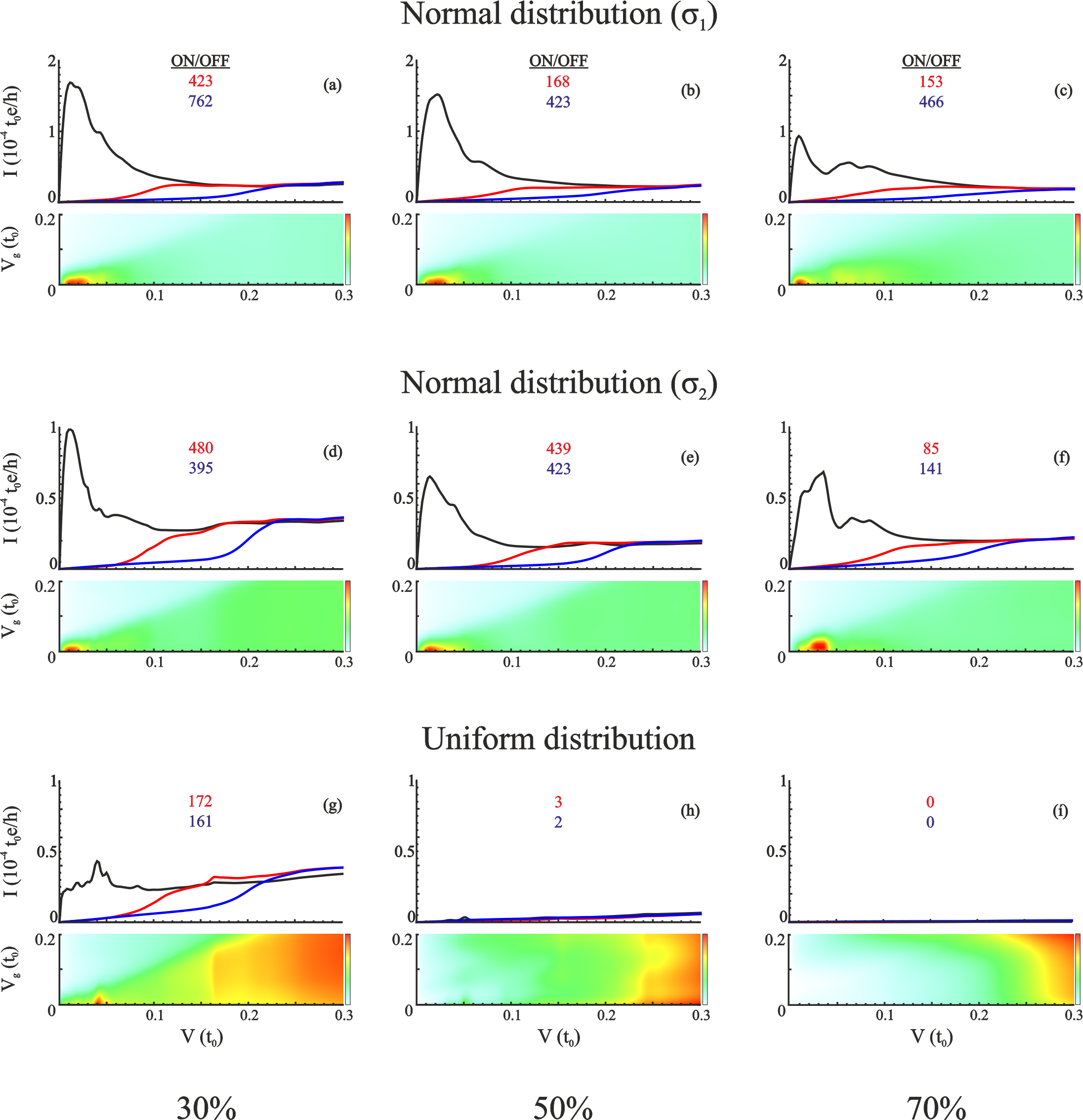}
\caption{Calculated I-V curves of the planar graphene TFET at different concentrations: 30\% (a,d,g), 50\% (b,e,h), and 70\% (c,f,i) for both normal distribution with dispersion parameters $\sigma_{1}=L/4$ and $\sigma_{2}=3L/4$ and uniform distribution. The gate voltage is taken to be $V_{g}=0$ (black), $0.1$ (red), and $0.2$ (blue). The ON/OFF ratios are shown. Colour maps show current densities normalized to the highest values for a range of $V$ and $Vg$.} 
\label{nu}
\end{figure*}

The original model of the planar graphene TFET considers the defect-free contacts~\cite{katkov}. 
In this case, the current-voltage (I-V) curves at room temperature and different gate voltages take the form shown in fig.~\ref{id}. According to our calculations, the main contribution to tunneling current comes from $t_{m}$, while $t_{1}$ gives a small correction.
As noted in~\cite{katkov}, the operation of the device is based on a possibility to manipulate the positions of peaks in the electronic density of states of zigzag graphene edges with relation to the energy window. 
Recently, we have shown that edge vacancies affect markedly the DOS at graphene edge: local DOS becomes redistributed between emerging subpeaks and the edge state which can even disappear under certain conditions~\cite{glebov}. 
In turn, it will degrade performance of graphene TFET.
This is clearly illustrated in fig.~\ref{dos}, where the DOS of graphene contacts is schematically shown together with the corresponding I-V curves at different gate voltages. 

For the defect-free graphene contact, the DOS has a high peak at the Fermi level associated with the edge state. 
The peaks are splitted due to the mutual influence of contacts. 
Applied bias voltage shifts the bands and induces an opening of the energy window thus allowing tunneling of charge carriers from filled to empty states (fig.~\ref{dos} a).
As a result, there appears a high peak in the I-V curve.
Further increase in bias voltage leads to a larger shift of the peaks so that opposite states have a reduced DOS (see fig.~\ref{dos} c). 
In this case, the tunnel current comes down and a region with negative differential resistance emerges.
Applied gate voltage provokes a shift of the Fermi level to the region of the low DOS in both contacts (fig.~\ref{dos} b) and, as a result, tunnel current turns out to be strongly suppressed until one of the DOS peaks meets the energy window (fig.~\ref{dos} c).

In the presence of edge vacancies, the local DOS was found to be changed in the neighborhood of the three nearest atoms to the vacancy~\cite{glebov}. 
This modification depends markedly on both the concentration and arrangements of the vacancies. 
Generally, two important features appear: (i) subpeaks near the Fermi level, and (ii) a decrease or even disappearance of the split central peak. 
Both factors influence the operation of the TFET. 
Namely, reduced peaks of edge states and the existence of subpeaks lead to a reduced tunnel current and peak smearing together with the appearance of additional subpeaks in the I-V curve (fig.~\ref{dos} d,f). 
Applied gate voltage moves the energy window to the region of DOS containing subpeaks. 
In this case, I-V curve shows non-zero current even at low bias voltages thus degrading the ability of the switching device (fig.~\ref{dos} e,f). 
It is clear that the TFET  will be completely destroyed when edge states disappear at both contacts. 

\section{Simulations}

We will consider edge vacancies of different (30, 50 and 70\%) concentrations located on both contacts with three types of distributions: normal, uniform, and periodic.
Defects are positioned by making use of a random number generator algorithm.
In the case of normal distribution, each atomic position is assigned a weighting factor 
\begin{equation}
f(x)=1/(\sigma\sqrt{2\pi})e^{-(x-\tilde{x})^{2}/(2\sigma^{2})},
\end{equation}
where $x$ denotes the number of the cell, and $\tilde{x}=L/2$ is the central cell which corresponds to the middle of the graphene ribbon. Dispersion parameters are taken to be $\sigma_{1}=L/4$ and $\sigma_{2}=3L/4$. 
In accordance with the weighting factor, the random number generator chooses $N=L\times n$ atomic positions in which the carbon atom is replaced by a vacancy with $n$ being the vacancy concentration.
For uniform distribution, each atomic position is assigned a weighting factor equal to 50\% chance of a defect appearance. In the case of a periodic distribution, the vacancy position is set manually.
Calculated tunneling current is averaged over ten random arrangements of the vacancies.

(i) \textit{Normal distribution}. This distribution is suitable for modeling situations when arrays of vacancies are mainly situated in the middle of the edge. 
The influence of defects leads to the appearance of subpeaks near the Fermi level while the central peak due to edge states decreases. 
At a concentration of 30\%,  we found no significant changes in the tunneling current (see fig. \ref{nu} a).
Increasing defect concentration causes an increase in the height and number of subpeaks as well as a further reduction of the central peak. 
This leads to a corresponding modification of the I-V curves (figs.~\ref{nu} b,c).

At low gate voltage, the energy window is shifted to the subpeak area near the Fermi level. 
This leads to a small current at a voltage less than $V_{g}$ and, taking also into account a decrease of the central peak, to the reduction of the ON/OFF ratio (figs.~\ref{nu} b,c). 
Further increase in defect concentration degrades performance of the TFET until the end of its operation at concentrations above 70\% due to the disappearance of edge states found in~\cite{glebov}. 
In the case of high dispersion ($\sigma_{2}=3/4L$) there appear additionally small groups of vacancies shifted to the edges of the sheet. 
This causes the formation of peaks in a wider energy interval, an increase in their height and a decrease in the central peak. 
As shown in figs.~\ref{nu} (d,e,f), with the increasing concentration of defects device efficiency markedly deteriorates. 

(ii) \textit{Uniform distribution}. In this case, contacts have a large number of single-located edge vacancies, whose mutual influence leads to a marked modification of the local DOS. Namely, the height of subpeaks is found to be greater than the split central peak, even at low concentrations of defects~\cite{glebov}. 
This is directly manifested in the I-V curves (see figs.~\ref{nu} (g,h,i)). 
The TFET performance is significantly reduced already at 30\% defect concentration and the transistor stops functioning at higher concentrations.   
 
\begin{figure}[!ht]
\includegraphics[scale=0.75]{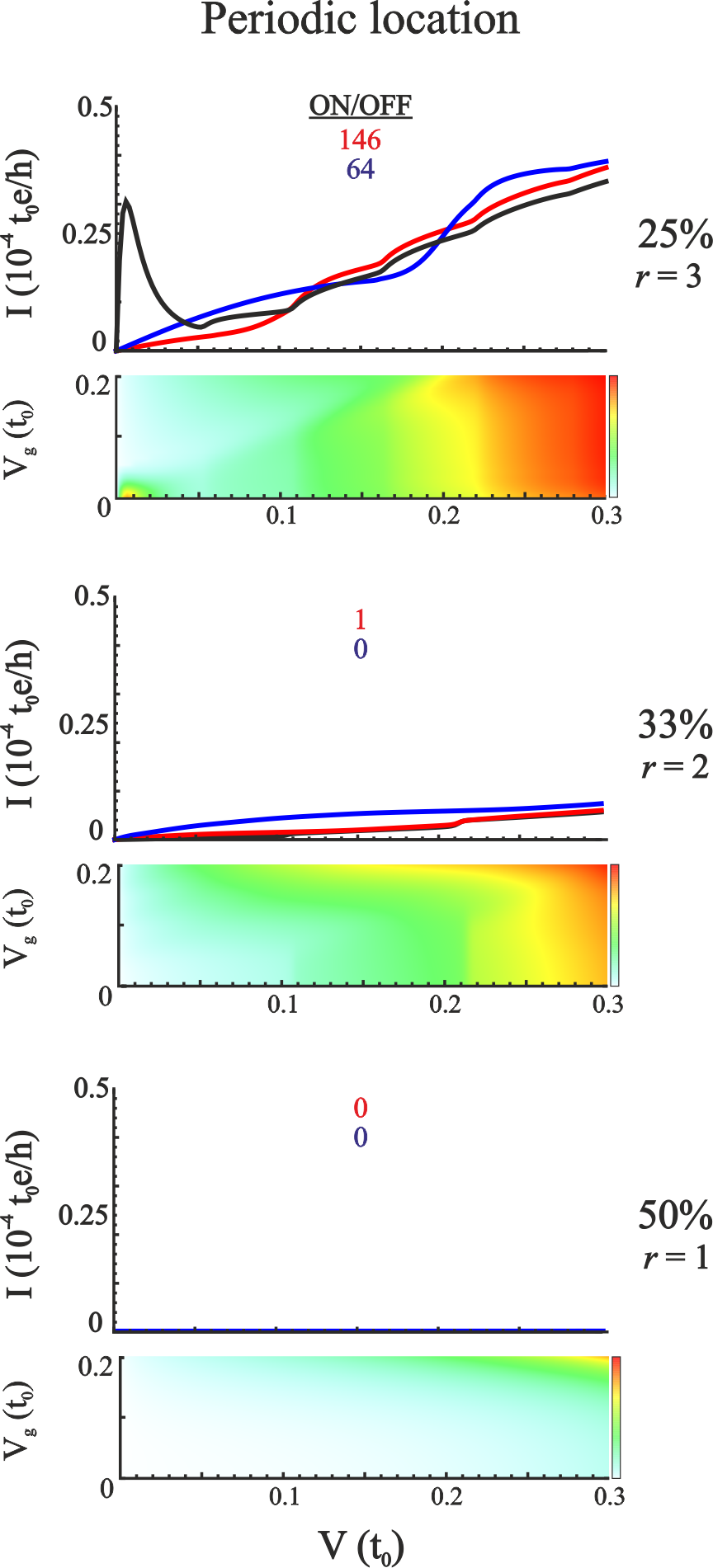}
\caption{Calculated I-V curves of the planar graphene TFET at $r=1$, $2$, and $3$ for  periodic distribution. The gate voltage is taken to be $V_{g}=0$ (black), $0.1$ (red), and $0.2$ (blue). The ON/OFF ratios are shown. Colour maps show current densities normalized to the highest values for a range of $V$ and $Vg$.}
\label{p}
\end{figure}

(iii) \textit{Periodic distribution}. We have considered three variants of distribution with edge vacancies located regularly with $r=1,2$ and $3$ atoms between them. 
It was found that the edge state disappears in all cases except for the location of single vacancies between three atoms because this concentration is too low to distort the total DOS~\cite{glebov}. 
This explains the calculated behavior of I-V curves shown in fig.~\ref{p}. 

\begin{figure}[!pht]
\includegraphics[scale=0.80]{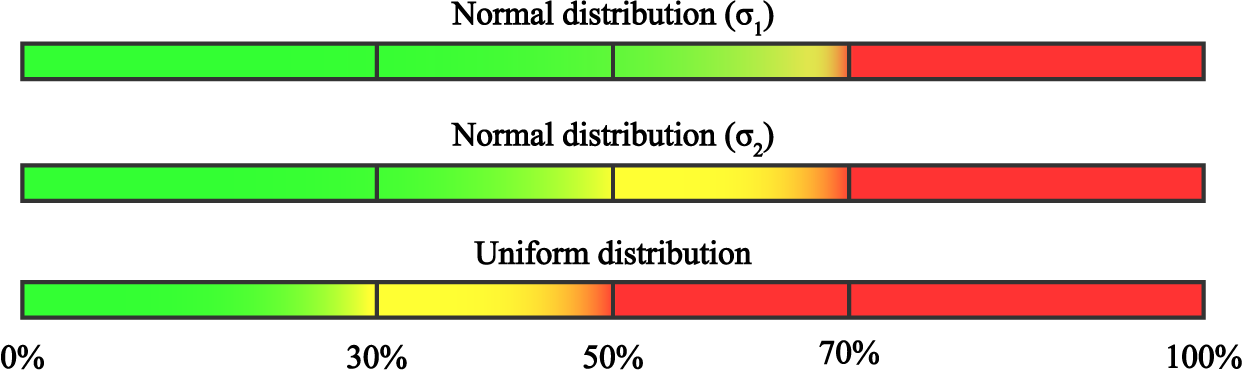}
\caption{Performance of the planar graphene TFET at different distributions and concentrations of edge vacancies: high performance (green), low performance (yellow), and the device does not work (red).} 
\label{con}
\end{figure}

In the last two cases, the device is not working. 
At $r=3$, the DOS is found to have a large number of subpeaks that fall in the energy window. 
This leads to a low ON/OFF ratio due to a high enough OFF current. 
At larger $r$, the defects do not have a noticeable effect on the DOS so that the device operates in much the same way as in the defect-free case. 

\section{Conclusion}

The main results of our calculations are illustrated in fig.~\ref{con}.
There are two factors that significantly affect the operation of the device. 
The first one is connected with the very existence of the edge states. 
Indeed, it is the presence of edge states that underlies the operation of the considered planar graphene TFET. 
We found that edge vacancies can destroy edge electronic states at some critical concentrations whose values depend on the type of distribution. 
This leads to negligible tunnel currents, as is directly shown in figs.~\ref{nu} (h,i) and~\ref{p}. 
It can be stated that the critical defect concentrations are above 70, 50, and 30 percent for normal, uniform, and periodic distribution, respectively. 
The second factor is the reduction of split central peak and the appearance of additional subpeaks in DOS as a result of mutual influence of edge vacancies. 
This reduces the ON and increases the OFF current thereby worsening the switching performance of the device.

Summing up, one can say that the more pronounced the central peaks in DOS due to edge states are, the better performance the device shows. 
Any factors leading to either reduction of the central peaks or emergency of additional subpeaks degrade device operation. 
Such changes may be caused by edge vacancies, as shown in this paper, and other reasons. 
This conclusion is of importance for the design of graphene-based nanoelectronic devices whose operation is based on the tunneling of charge carriers through the edge states.

\end{document}